\documentclass[10pt]{iopart}

\usepackage{graphicx}
\usepackage[colorlinks=true,%
            linkcolor=blue,%
            urlcolor=blue,%
            citecolor=blue,%
            filecolor=blue,%
            bookmarksopen=true,%
            pdfauthor={J. P. Ramos-Andrade},%
            pdftitle={AB_QD_MBS},%
            pdfsubject={AB_QD_MBS_Manuscript},%
            pdfpagemode=UseOutlines]{hyperref}

\usepackage{amssymb}

\newcommand{\ve}{\varepsilon}

\begin{document}
\title{Bound states in the continuum poisoned by Majorana fermions}
\author{D Zambrano$^1$, J P Ramos-Andrade$^{1}$, P A Orellana$^1$}
\address{$^1$ Departamento de F\'isica, Universidad T\'ecnica Federico Santa Mar\'ia, Casilla 110 V, Valpara\'iso, Chile}
\ead{david.zambrano@usm.cl}
\begin{abstract}
In this work, we study the bound states in the continuum (BICs) in a system formed by a triple quantum dot array embedded between two one-dimensional topological superconductors, both hosting Majorana bound states (MBSs) at its ends. The results show the formation of BICs with topological characteristics due to the presence of MBSs. The latter is a consequence of the interplay between the BIC arising from quantum dots states by means of energy level symmetry breaking through gate voltages, and MBSs leaked into the quantum dots. The BIC is not observed when both TSCs are in long wire limit, i. e. for vanishing inter MBSs coupling, while it projects into the electronic transmission whenever the inter MBSs couplings are away from zero, regardless if they are different and/or the phase difference between both TSCs.
We study the behavior of BICs poisoned by MBSs as a function of the parameters that are controlling the system. We believe our findings could be useful to implement a protection tool for BICs using MBSs based on tunable gate voltages.
\end{abstract}
\maketitle
\ioptwocol

\section{Introduction}

Bound states in the continuum (BICs) were predicted by von Neumann and Wigner at the dawn of quantum mechanics \cite{vonNeumann1929}. They found that states embedded in the continuum band keep bounded for a particular kind of oscillating potential. Recently, interest in BICs investigation has been increasing due to the observation of these types of states in photonic systems. Since interference phenomena takes place in electronic systems in analogy with the photonic ones, the inherent possibility of the presence of BICs emerges \cite{Hsu2016, Ramos2014,Cortes2014,Gonzalez2013,Gonzalez2010,DeGuevara2006}.

On the other hand, the Majorana fermions (MFs) were first predicted by Ettore Majorana in 1937, which are fermionic particles that are their own antiparticles \cite{Majorana1937, Wilczek2009, Franz2010}. Since then, there have been several efforts to discover this kind of particle. A natural MF candidate is the neutrino, however, up to now, there is no evidence to support it. In the context of condensed matter physics, Kitaev predicted the realization of localized MFs, Majorana bound states (MBSs), at the ends of a semiconductor-superconductor nanowire in topological phase, the so-called Kitaev chain \cite{Kitaev2001, Kitaev2003}. Nowadays, different experiments suggest that this theoretical prediction has been confirmed by means of physical realization of the Kitaev proposal \cite{Mourik2012,Deng2012,Das2012,Lee2012,Finck2013,Churchill2013}. Recently, the investigation of MBSs has had a great deal of attention as these states are seen to be useful in quantum computation implementations since they satisfy non-Abelian statistics and can be manipulated with braiding operations \cite{Alicea2011, Nayak2008, Leijnse2011, Bravyi2000, Kitaev2001, Kitaev2003, Pachos2012, Kraus2013, Albrecht2016, Beenakker2013, Laflamme2014,hoffman2016universal}. In quantum dot (QD) systems, a special signature of the presence of MBS was established as a half-integer conductance at zero energy \cite{Liu2011} when the MBS is side-coupled with the QD. Later, Vernek \textit{et al}. \cite{Vernek2014} have shown that this zero-bias anomaly is due to MBS leaking into the QD, and it is robustly pinned against changes in QD energy level, which has recently been verified \cite{Deng2016}. In this scenario, a proposal of Majorana-based qubit readout technology was carried out using BICs mechanisms in an embedded QD between topological superconductors (TSCs) \cite{Ricco2016} and \cite{hoffman2016universal}, and recently a theoretical encryption device based on BICs is available in a TSC coupled to a double QD structure \cite{Guessi2017}, also a spin-dependent coupling between QDs and topological quantum wires has been proposed \cite{hoffman2017spin}.

Multiple QDs systems have the possibility to achieve BICs arising naturally from interference phenomena. However, there is no information on how this kind of BIC behaves in interplay with leaking MBSs. In this work, we address the latter and calculate the formation of BICs poisoned by MBS in a triple QD system with two side-coupled topological superconductors (TSCs) nanowires. MBS have topological characteristics, on the other hand, there are BICs arising from the hybridization of the isolated triple QD array. Then, both kinds of bound states coexist in our system. Our results show the buildup of BICs with topological characteristics due to the presence of the MBSs. For symmetric QDs levels, the presence of two BICs has a projection in the electronic transmission as two side half-maximum resonances, while by breaking the QDs symmetry through a gate voltage, another BIC projection in transmission is robustly pinned at zero energy against this symmetry breaking, where its amplitude depends on the MBSs overlapping parameter.

The behavior of the MBS inheriting their topological properties to the BICs states can be seen by looking the zero energy states, which when are strictly degenerated, these MBS are hiding the BIC. We show that by manipulating the MBS \cite{Alicea2011} splitting, the BIC can be unveiled/hidden.

This paper is organized as follows: Section \ref{model} presents the system Hamiltonian and method used to obtain quantities of interest; Section \ref{results} shows the results and the corresponding discussion, and finally, the concluding remarks are presented in Section \ref{summary}.

\section{Model}\label{model}

\begin{figure}[!t]
\centering
\includegraphics[width=0.9\linewidth]{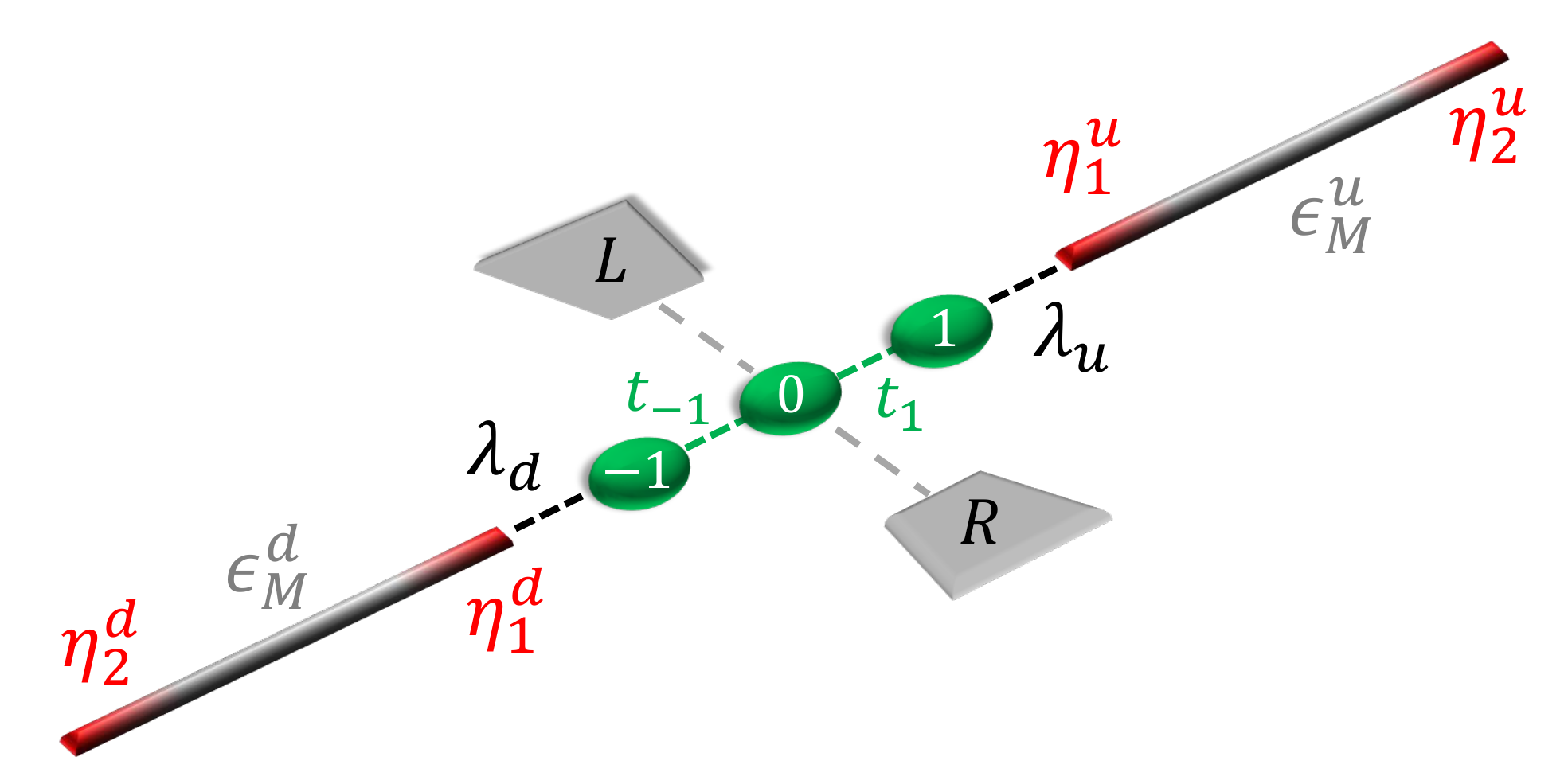}
\caption{Model setup: Crossbar-shaped QD-TSCs system. A triple QD array (green) coupled to two normal leads, labeled as L and R (solid gray), and two TSCs (gray tones) $u$ and $d$, each hosting two MBSs (red), $\eta_{1}^{u(d)}$ and $\eta_{2}^{u(d)}$.}
\label{fig.montaje}
\end{figure}

The system under study consider a crossbar-shaped form by a linear array of three QDs, two normal leads and two TSCs hosting MBSs at its edges. The central QD (QD$_{0}$) is connected with both leads. Each QD located at the end of the array has a side-coupled TSC, as shown in Fig.\ \ref{fig.montaje}.

We model the system with an effective low-energy Hamiltonian, which has the form
$H = H_{\scriptsize{\textrm{leads}}} + H_{\scriptsize{\textrm{dots}}} + H_{\scriptsize{\textrm{dot-leads}}} + H_{\scriptsize{\textrm{dot-M}}} + H_{\scriptsize{\textrm{M}}}$, where the first three terms on the right side correspond to the regular electronic contribution, given by
\begin{eqnarray}
H_{\scriptsize{\textrm{leads}}} &=& \sum_{\alpha,{\bf k}}{ \varepsilon_{\alpha, {\bf k}} c_{\alpha, {\bf k}}^\dagger c_{\alpha,{\bf k}} }\,,
\\
H_{\scriptsize{\textrm{dots}}} &=& \sum_{j=-1}^{1}{ \varepsilon_{j} d_{j}^\dagger d_{j}} + \sum_{j=-1}^{0} {t\,d_{j}^\dagger d_{j+1} }+\textrm{H.c.}\,,\\
H_{\scriptsize{\textrm{dot-leads}}} &=& \sum_{\alpha,{\bf k}}{ V_{\alpha} d_{0}^\dag c_{\alpha, {\bf k}}} + \textrm{H.c.} \,,
\end{eqnarray}
where $c_{\alpha, {\bf k}}^\dag ( c_{\alpha, {\bf k}} )$ is the electron creation (annihilation) operator with momentum ${\bf k}$ and energy $\varepsilon_{\alpha, {\bf k}}$ in the lead $\alpha = \textrm{L,R}$. $d_{j}^\dagger (d_{j})$ is the electron creation (annihilation) operator in the $j$th QD, with single energy level $\varepsilon_{j}$. $t$ is the inter-dot coupling and $V_{\alpha}$ is tunneling hopping between the lead $\alpha$ and QD$_{0}$.

The two last terms in the Hamiltonian, $H_{\scriptsize{\textrm{M}}}$ and $H_{\scriptsize{\textrm{dot-M}}}$, correspond to MBSs and their couplings with QDs, respectively. They are given by
\begin{eqnarray}
H_{\scriptsize{\textrm{dot-M}}} &=& \left( \lambda_d d_{_{-1}} - \lambda_d^* d{^\dagger_{_{-1}}} \right) \eta_{1}^d + \left( \lambda_u d_{_{1}} - \lambda_u^* d{^\dagger_{_{1}}} \right) \eta_{1}^u ,\label{HdotM}\\
H_{\scriptsize{\textrm{M}}} &=& i\varepsilon_{M}^d\eta_{1}^d\eta_{2}^d + i\varepsilon_{M}^u\eta_{1}^u\eta_{2}^u\label{HM}\,,
\end{eqnarray}
where $\eta_{\beta}^{l}$ denotes the MBS operator, which satisfies both $\eta_{\beta}^{l}=[\eta_{\beta}^{l}]^{\dag}$ and $\{\eta_{\beta}^{l},\eta_{\beta'}^{l'}\}=\delta_{\beta,\beta'}\delta_{l,l'}$, being $\beta=1,2$ and $l=d,u$. In addition, $\lambda_{d(u)}$ is the tunneling coupling between $\eta_{1}^{d(u)}$ and the QD$_{-1(1)}$, and $\epsilon_{_{M}}^{l} \propto \exp(-L_{l}/\zeta)$ is the coupling strength between two MBSs in the same TSC, where $L_{l}$ denotes the wire length and  $\zeta$ is the superconducting coherence length. Without loss of generality, we fixed $\lambda_{d}=\lambda_{d}^{\ast}=|\lambda_{d}|$ y $\lambda_{u}=|\lambda_{u}|\exp[i\theta/2]$, where $\theta$ is the phase difference between both TSCs.

\begin{figure}[!t]
\centering
\includegraphics[width=1.0\linewidth]{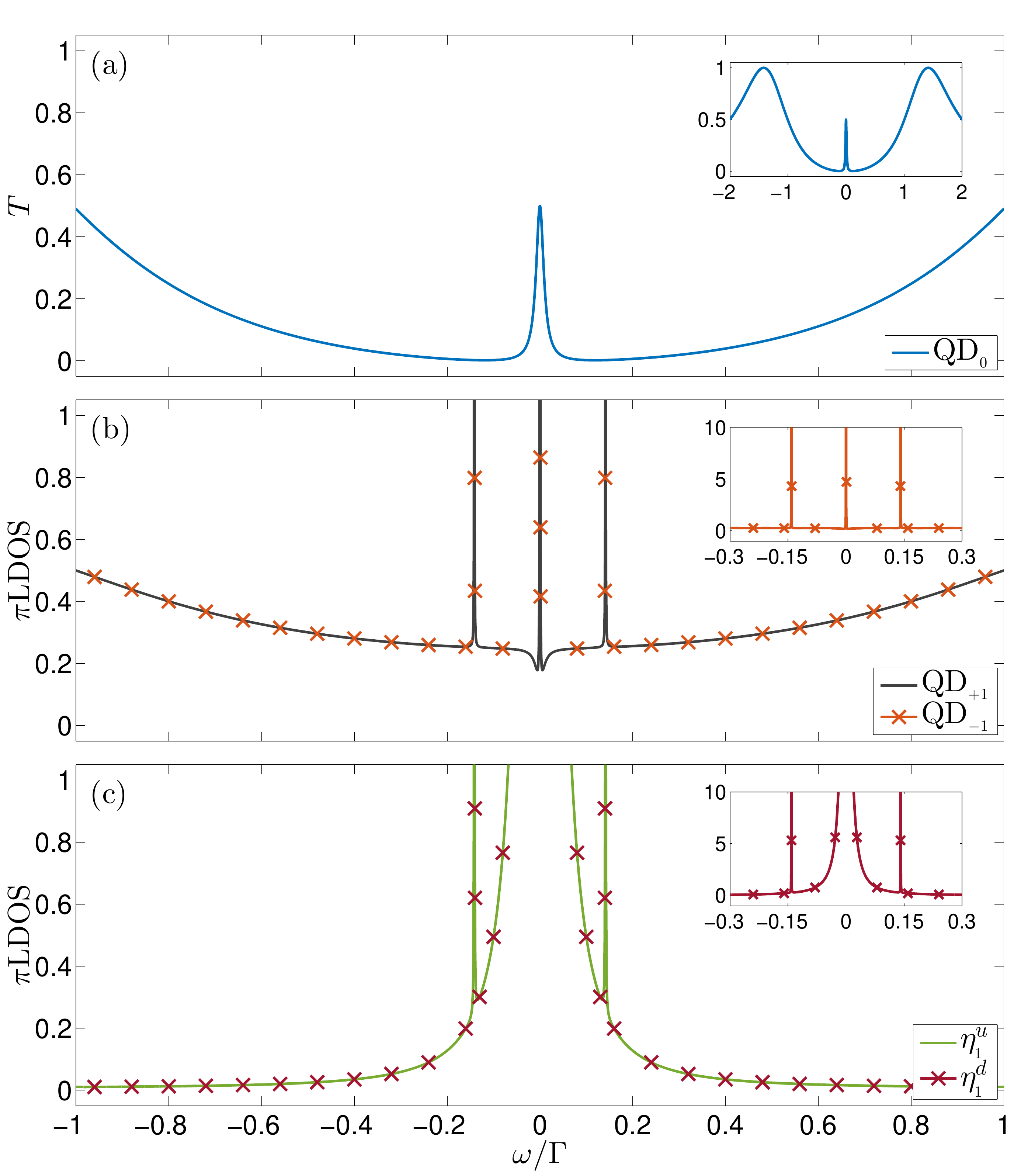}
\caption{Transmission and LDOS as a function of energy through the central QD$_{0}$ in the long wire limit ($\epsilon_{M}^{u(d)} = 0$) with $\varepsilon_{0} = \varepsilon_{\pm1} = 0$. Panel (a) shows the transmission as a function of the energy $\omega$ which is qualitatively equivalent to the LDOS in the central QD. Panel (b) shows the LDOS for the two external QDs (QD$_{\pm 1}$). Panel (c) shows LDOS for the Majorana modes $\eta_1^{u(d)}$ which are couple with the external QDs}
\label{fig.one_panel}
\end{figure}

A useful way to treat the system analytically is by writing each MBS as a superposition of regular fermionic operators as $\eta_{1}^{l}=(f_{l}+f_{l}^{\dag})/\sqrt{2}$ and $\eta_{2}^{l}=-i(f_{l}-f_{l}^{\dag})/\sqrt{2}$
which satisfy both $\{f_{l},f_{l'}\}=\{f_{l}^{\dag},f_{l'}^{\dag}\}=0$ and $\{f_{l},f_{l'}^{\dag}\}=\delta_{l,l'}$. Then, Eqs.\ (\ref{HdotM}) and (\ref{HM}) transform to
\begin{eqnarray}
H_{\scriptsize{\textrm{dot-M}}} &=& \frac{1}{\sqrt{2}}\left(\lambda_d d_{_{-1}} - \lambda_d^* d{^\dagger_{_{-1}}} \right) \left(f_{d}+f_{d}^{\dag}\right)\nonumber\\ &+& \frac{1}{\sqrt{2}}\left(\lambda_u d_{_{1}} - \lambda_u^* d{^\dagger_{_{1}}} \right) \left(f_{u}+f_{u}^{\dag}\right)\,,
\label{HdotMf}\\
H_{\scriptsize{\textrm{M}}} &=& \epsilon_{M}^d\left(f_{d}^{\dag}f_{d}-\frac{1}{2}\right) + \epsilon_{M}^u\left(f_{u}^{\dag}f_{u}-\frac{1}{2}\right)\label{HMf}\,.
\end{eqnarray}

The main contribution of the leads is to include a self-energy $\Sigma_{\alpha}^{e(h)}$ for electrons(holes). In the wide-band limit approximation it is energy-independent, such as $\Sigma_{\alpha}^{e(h)}\equiv-i\Gamma_{\alpha}^{e(h)}$, and it fulfills electron hole symmetry, hence $\Gamma_{\alpha}^{e(h)}\equiv\Gamma_{\alpha}$. We consider symmetric QD$_{0}$-leads coupling $\Gamma_{\alpha}\equiv\Gamma/2$, so $\Gamma_{\scriptsize{\textrm{L}}}+\Gamma_{\scriptsize{\textrm{R}}}=\Gamma$. In this scenario the transmission probability can be written as $T(\omega)=-\Gamma\,\textrm{Im}\left[G_{0}^r(\omega)\right]$, with $G_{0}^r$ being the QD$_0$ retarded Green function. Finally, the LDOS can be written as $\textrm{LDOS}_{j}(\omega) = -(1/\pi)\textrm{Im}\left[G_{j}^r(\omega)\right]$ for QDs, and $\textrm{LDOS}_{\beta,l}(\omega) = -(1/\pi)\textrm{Im}\left[G_{{\beta,l}}^r(\omega)\right]$ for MBSs. Throughout the next sections, we focus on these quantities to explore the coexistence and interplay of BIC wtih MBSs.

\section{Results}\label{results}

\subsection{Without phase difference, $\theta=0$}\label{theta0}

Before going into numerical results, an analysis of the eigenvalues is performed for the non-leads Hamiltonians of the proposed system appearing in Fig.\ \ref{fig.montaje} for the case without phase difference. These are closely related to the full system Green's function poles, and give reliable information about energy localization of the states. For a particular case, assuming $|\lambda_{u(d)}| = \lambda$, $\epsilon_{_M}^{u(d)} = \epsilon_{_M}$, $\varepsilon_{_0} = 0$, $\varepsilon_{_{\pm 1}} = \pm\Delta$ and $t_{_1} = t_{_{-1}} = t$ the eigenvalues can be written out as follows:
\begin{eqnarray}
	&&\omega_{_1} = 0\,,\label{w1}\\
	2\left[\omega_{_2}^{i,j}\right]^{2} &=&  \epsilon_{_M}^2 + \Delta^2 + 2\left( \lambda^2 + t^2 \right)\nonumber  \\ &\pm&  ( \left( \epsilon_{_M}^2 + \Delta^2 \right)\left[ \epsilon_{_M}^2 + \Delta^2 + 4\left( \lambda^2 + t^2 \right) \right]\nonumber \\ &-& 4\epsilon_{_M}^2\left( \Delta^2 + 2t^2 \right) + 4\left( \lambda^2 \pm t^2 \right)^2 )^{1/2}   \,, \label{w2}
\end{eqnarray}
where the energy $\omega_{_1}=\epsilon_{_0}=0$ has double degeneracy, while the other eight energies are contained in $\omega_{_2}$, each corresponding to a particular combination of $+$ and $-$ signs. For a fixed $\Delta = 0$, the following energies are computed from Eq. (\ref{w2}) breakdown:
\begin{eqnarray}
    &~&\omega_{_2}^{+,-}(\Delta = 0)= \pm t\sqrt{2}\,, \label{w21}\\
    &~&\omega_{_2}^{-,-}(\Delta = 0) = \pm \sqrt{ \varepsilon_{_M}^2 + 2\lambda^2 }\,, \label{w22}
\end{eqnarray}
\begin{eqnarray}
    2\left[\omega_{_2}^{\pm,+}(\Delta = 0)\right]^{2} &=& \epsilon_{_M}^2 + 2\left( \lambda^2 + t^2 \right) \\
    &\pm&  \sqrt{  \left[ \epsilon_{_M}^2 + 2\left( \lambda^2 + t^2 \right) \right]^2 - 8\epsilon_{_M}^2 t^2 } \nonumber \,,\label{w23}
\end{eqnarray}
where it can be noted that if the MBSs overlapping vanishes, $\epsilon_{_M}=0$, two of the four possibles values of Eq.\ (\ref{w23}) take the zero value.

In what follows, all the numerical results are performed at temperature $\mathcal{T} = 0$, in which the conductance is $G(\varepsilon_{\scriptsize{\textrm{F}}})=(e^{2}/h)T(\omega=\varepsilon_{\scriptsize{\textrm{F}}})$, with $\varepsilon_{\scriptsize{\textrm{F}}}$ being the Fermi energy, and all the energy parameters are given in units of fixed $\Gamma=1$ meV. Throughout this work, the inter dot couplings are fixed at $t_{j} = \Gamma$ and a weak QD-TSC coupling $|\lambda_{u(d)}|=\lambda=0.1\,\Gamma$ is also considered.

\begin{figure}[!b]
\centering
\includegraphics[width=1.0\linewidth]{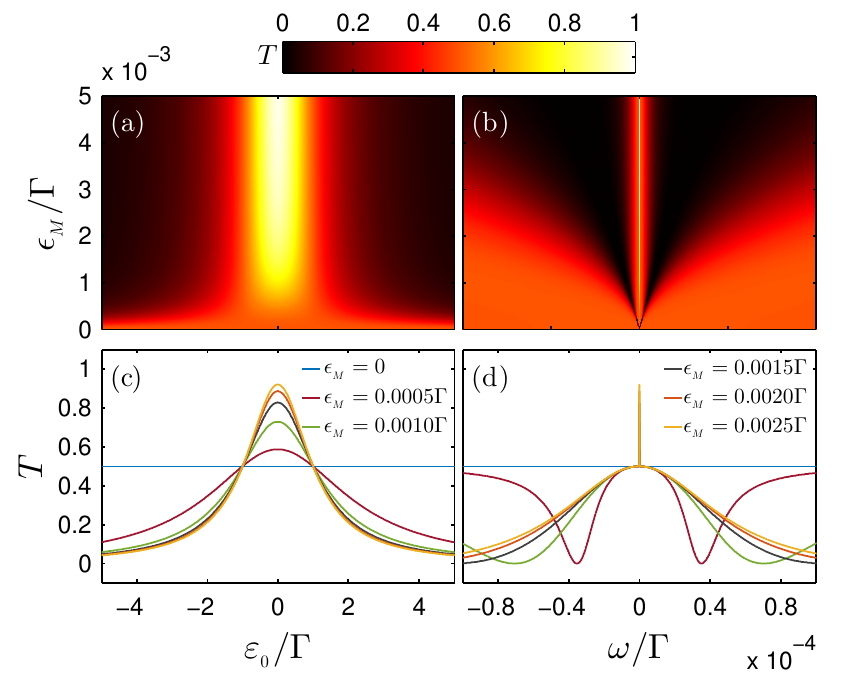}
\caption{Transmission $T$ contour plot as a function of the inter Majorana coupling $\epsilon_{_M}$, by changing (a) the central QD level $\varepsilon_{_0}$ and (b) energy $\omega$. Panels (c) and (d) correspond to horizontal cuts in panels (a) and (b), respectively. In all panels $\Delta = 0.01\Gamma$.}
\label{fig.four_panel}
\end{figure}

First we display LDOS for all the elements in the system and transmission across QD$_{0}$ in Fig.\ \ref{fig.one_panel}, in which we use a long wire limit, i. e. $\epsilon_{M}^{u(d)}=0$. The transmission shows two broad lateral maximums [inset in panel (a)] due to the QDs hybridization [see for instance Eq.\ (\ref{w21})] and it is clear that the zero energy half-integer is still observed, solid blue lines in panel (a). Interestingly, this behavior is evidence that the leaking of the two MBSs into the central QD occurs even through the two lateral QDs. The LDOS for the MBSs $\eta_{1}^{u(d)}$ and the lateral QD$_{\pm 1}$ show two symmetric BICs placed at energies $\pm\sqrt{2}\lambda$, which are given by Eq.\ (\ref{w22}). These states correspond to hybridized MBSs, which do not show apparent projection in the transmission.

\begin{figure}[!t]
\centering
\includegraphics[width=1.0\linewidth]{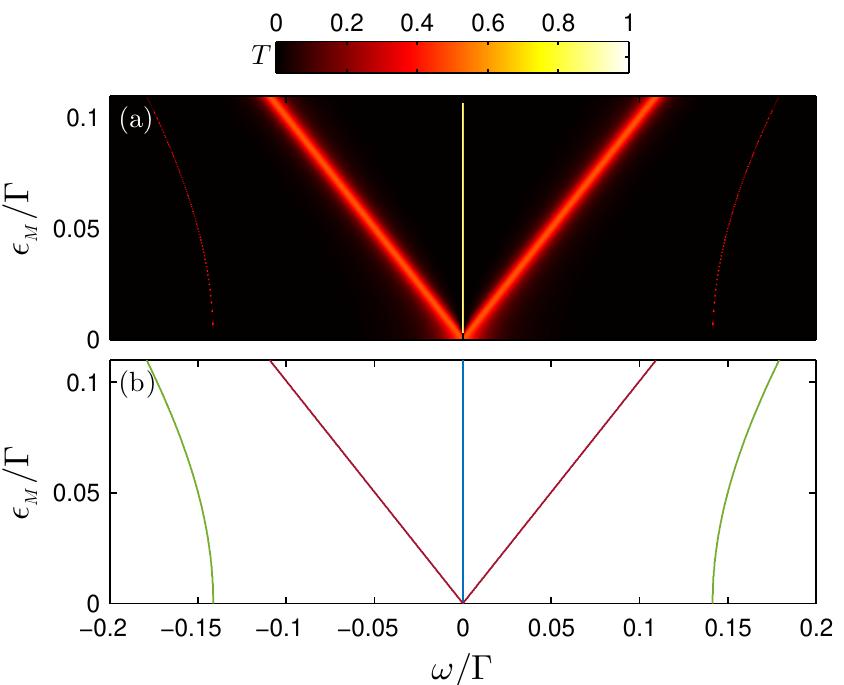}
\caption{(a) Transmission $T$ contour plot as function of $\epsilon_{_M}$ and $\omega$, basically a zoom out of Fig.\ \ref{fig.four_panel}(b). (b) Eigenvalues of the disconnected system given by Eq.\ (\ref{w1}) (blue line), Eq.\ (\ref{w22}) (red lines) and Eq. (\ref{w23}) (green lines) using the minus sign inside the square root, as a function of $\epsilon_{_M}$. In both panels we fixed $\varepsilon_{_0} = 0$ and $\Delta=0$. The eigenvalues not shown are located far from zero ($\omega\sim\sqrt{2}\,\Gamma$), out of our scope, and are given by Eq.\ (\ref{w21}) and Eq.\ (\ref{w23}) considering the plus sign inside the square root.}
\label{fig.two_panel}
\end{figure}

To explore the above behavior in more detail, we introduce a small asymmetry energy parameter by setting the gate voltages of the lateral QDs as $\varepsilon_{{\pm1}} = \pm\Delta$. For the isolated QDs case ($\lambda=0$), this allows revealing a BIC in the transmission coming from the hybridization of QDs states. In our case ($\lambda\neq 0$) we use it to study the interplay of the different phenomena.

Figure\ \ref{fig.four_panel} displays the transmission shape going out from the long-wire limit, considering $\epsilon_{_M}^{u(d)}=\epsilon_{_M}$, i. e. symmetrical MBSs overlapping. In Figs.\ \ref{fig.four_panel}(a) and \ref{fig.four_panel}(c) the cases with fixed zero-energy ($\omega=0$) are considered. As $\epsilon_{_M}$ is turned on, the transmission around $\varepsilon_{_0}=0$ begins to increase continuously from the half-integer conductance at zero energy to the unitary limit.

In Figs.\ \ref{fig.four_panel}(b) and \ref{fig.four_panel}(d) we explore the transmission considering $\varepsilon_{_0} = 0$. In this case, by moving away from $\epsilon_{_M}=0$ the transmission begins to unveil states placed at energies given by the Eqs.\ (\ref{w1}) and (\ref{w2}). It is important to note that these states do not appear in the transmission for the fully symmetric case ($\Delta = 0$), as is shown in Fig.\ \ref{fig.one_panel}. This behavior of the transmission represents a signature of the existence of a hidden BIC, covered by the MBSs leakage when $\epsilon_{_M}=0$ and being gradually exposed as $\epsilon_{_M}$ increases.

Furthermore, in Fig.\ \ref{fig.two_panel} we compare the behavior of the transmission with the eigenvalues of the disconnected system. Figure\ \ref{fig.two_panel}(a), which is a zoom out of Fig.\ \ref{fig.four_panel}(b), displays the evolution of transmission shape with $\epsilon_{_M}$, while Fig.\ \ref{fig.two_panel}(b) shows the eigenvalues using $\Delta=0$ ($\varepsilon_{_{\pm1}} = 0$) given by Eqs.\ (\ref{w1}), (\ref{w22}) and (\ref{w23}). We focus in a region near to zero energy, where only those eigenvalues that can be attributed to BICs were considered. Therefore, Fig.\ \ref{fig.two_panel}(b) shows the behavior of six of these states as a function of $\epsilon_{_M}$, and the other four states occur at energies far from zero, at $\sqrt{2}\Gamma$. With a small value of $\Delta$, i. e. $\Delta^{2}/\Gamma^{2}\ll 1$, the numeric resemblance is remarkable, and the relation between transmission resonances and eigenvalues allow us to predict exactly for which energy a state can be accessed due to its projection on the transmission curve.

\begin{figure}[!t]
\centering
\includegraphics[width=1.0\linewidth]{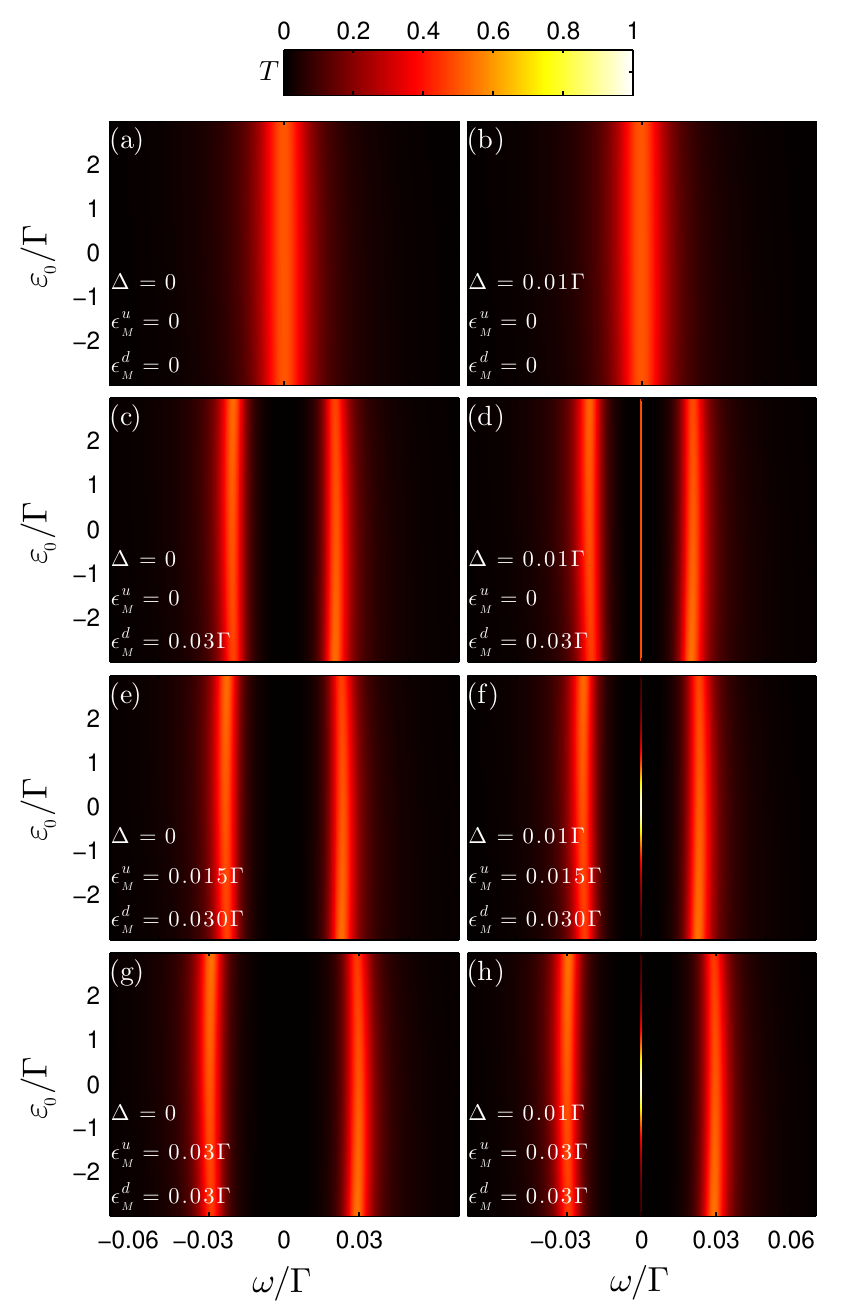}
\caption{Transmission $T$ contour plot as a function of the energy $\omega$ and central QD energy level $\varepsilon_{_0}$ for different values of $\Delta$ and $\epsilon_{_M}^{u(d)}$.}
\label{fig.eight_panel}
\end{figure}

It is important to highlight how some of these states can be suppressed/covered by the influence of the MBSs leakage. This poisoning behavior only allows protection of the BICs when the TSCs nanowires are long enough to reach $\epsilon_{M} = 0$. A way to manipulate $\epsilon_{M}$ is by using a \textit{keyboard} of locally tunable gates beneath the TCS nanowire \cite{Alicea2011}, where the effective topological wire length can be manipulated by transporting the MBS in the free end of the TSC (for instance $\eta_{2}^{u(d)}$) to a position closer/further from the fixed MBS, coupled to the QD ($\eta_{1}^{u(d)}$). This setup will allow a measurement of the poisoned BICs.

From this point we addressed the robustness of this behavior for non-symmetric TSCs. In order to give a further analysis of the system, in Fig.\ \ref{fig.eight_panel} a transmission color map regarding $\varepsilon_{_0}$ and $\omega$ is performed, where we start from the $\epsilon_{M}^{u(d)}=0$ case to finish in $\epsilon_{M}^{u(d)} =0.3\,\Gamma$, passing through different $\epsilon_{_M}^u \neq \epsilon_{M}^d$. In Fig.\ \ref{fig.eight_panel} left and right panels we fixed $\Delta = 0$ and $\Delta = 0.01\Gamma$, respectively. As is expected, in long-wire limit $\epsilon_{M}^{u(d)}=0$ a robust zero energy half-integer transmission is shown Fig.\ \ref{fig.eight_panel}(a) and Fig.\ \ref{fig.eight_panel}(b). On the other hand, from Figs.\ \ref{fig.eight_panel}(c) and \ref{fig.eight_panel}(d), if one of the TSC leaves the long wire limit ($\epsilon_{M}^d = 0.03\,\Gamma$), the half-integer transmission splits, being placed at $\omega\sim\pm\epsilon_{M}^{d}$ (due to the complexity of the eigenvalues expressions the exact value must be computed numerically) regardless $\Delta$. At zero energy ($\omega=0$) the transmission becomes strongly dependent of $\Delta$. For $\Delta=0$, the expected zero energy resonance is entirely suppressed, being blocked by the corresponding side coupled QD, while $\Delta\neq 0$ restores the zero energy half-integer resonance. As a consequence, at this point the BIC is still poisoned by the leaked MBS from the TSC that remains in the long-wire limit ($\epsilon_{_M}^u = 0$). In the next four panels, Figs.\ \ref{fig.eight_panel}(e)-(h), both TSCs leave the long wire limit. In the case of panels (e) and (f) we consider $\epsilon_{_M}^{d}/\epsilon_{_M}^{u}=2$, for which the two lateral QDs keep a similar behavior (regardless $\Delta$). At zero energy for $\Delta=0$ the corresponding state does not show any projection in transmission, while for $\Delta\neq0$ the MBS leakage and the BIC can interplay, being a poisoned BIC. At this point it reaches an integer transmission around $\varepsilon_{0}=0$, but it is worth mentioning that its emergence is continuous, as we showed before. Finally, in panels (g) and (h), when both TSCs reach the same wire length ($\epsilon_{_M}^{u(d)} = \epsilon_{_M} = 0.03\,\Gamma$) these two states are placed at $\omega=\pm \epsilon_{_M}$, and this value can be obtained from Eq.\ (\ref{w2}) or in an approximated way by Eq.\ (\ref{w23}). Just like the case discussed above, the BICs remain robust around $\varepsilon_{_0} = \omega = 0$. Therefore, we conclude that the non-topological BIC arising with $\Delta\neq0$ is poisoned whenever at least one of the TSCs keeps the long wire limit, i.e. $\epsilon_{_M}^{u}=0$ and/or $\epsilon_{_M}^{d}=0$.

\subsection{General phase difference}

\begin{figure}[!t]
\centering
\includegraphics[width=1.0\linewidth]{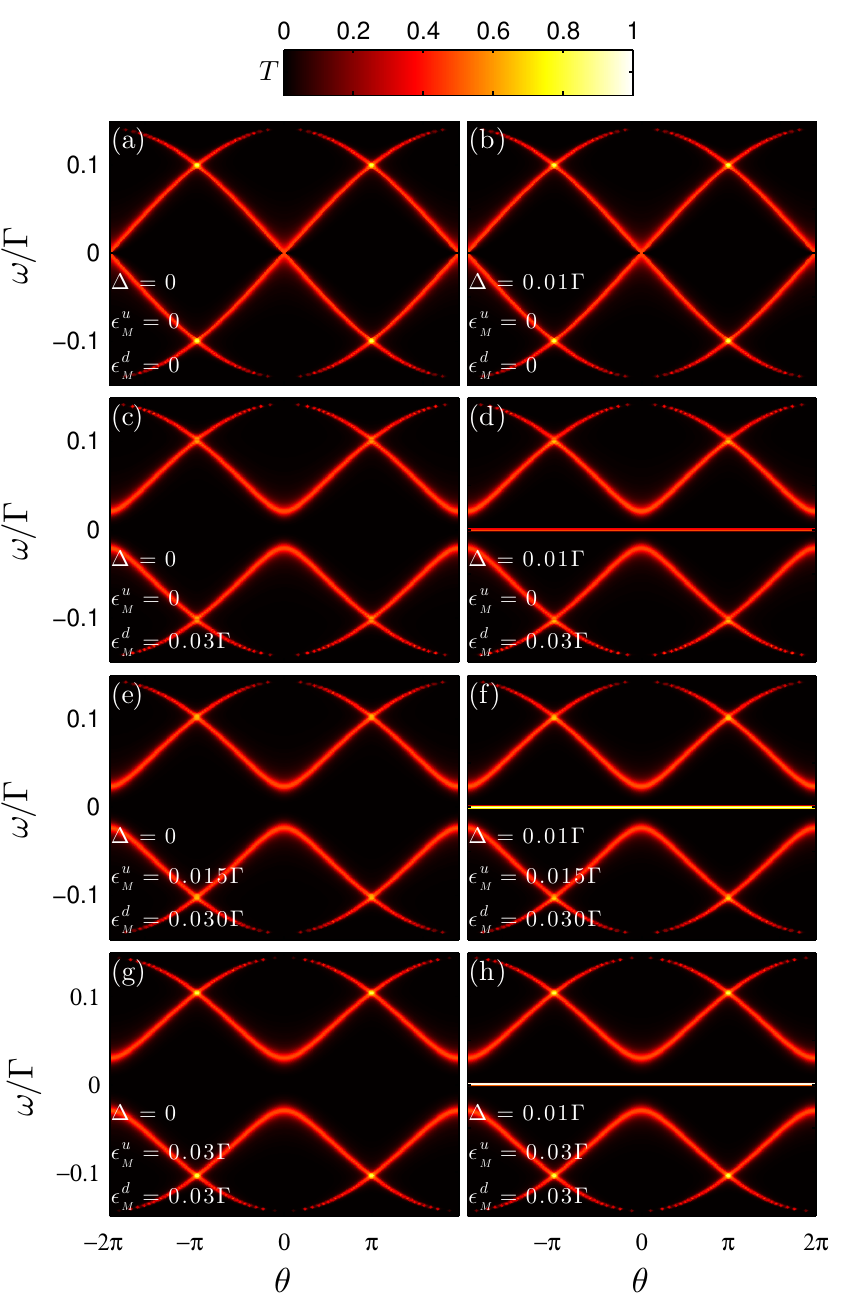}
\caption{Transmission $T$ contour plot as function of the phase difference $\theta$ and the energy $\omega$ for different values of $\Delta$ and $\epsilon_{_M}^{u(d)}$. $\ve_{0}=0$.}
\label{fig6}
\end{figure}

In orden to give a complete analysis of our system, we plot in Fig.\ \ref{fig6} a transmission color map as function of phase difference and energy. The case with $\epsilon_{M}^{d(u)}=0$ is displayed in Figs.\ \ref{fig6}(a) and \ref{fig6}(b), using $\Delta=0$ and $\Delta=0.01\Gamma$, respectively. We find that the bound states are completely suppressed around $\omega=0$, since MBSs interact destructively whenever TSCs are out of phase, i. e. $\theta\neq 2\pi m$, where $m$ is an integer number. The discussion above given in subsection \ref{theta0} can be extended for $\theta=2\pi m$.

On the other hand, Figs.\ \ref{fig6}(c), (e) and (g), with $\Delta=0$, shows similar shape around $\omega=0$, it is no projection of any state in the transmission, regardless the phase difference. However, the behavior changes as we consider $\Delta\neq 0$. In Fig.\ \ref{fwig6}(d), where only one TSC is within long wire limit ($\epsilon_{M}^{u}=0$), a half-integer transmission is observed at $\omega=0$, as a consequence of the bound states present in the system, being the BIC covered by the MBS leakage regardless the phase difference $\theta$. The BIC unveiling is achieved regardless $\theta$ whenever both TSCs have non-vanishing inter MBSs coupling ($\epsilon_{M}^{d(u)}$), as we show in Figs.\ \ref{fig6}(f) and (h).

It is important to highlight that the switching for hiding/unveiling a BIC in transmission does not depend on the phase difference between the TSCs and, for fixed $\Delta\neq 0$, can be controlled just by tuning the topological length of TSCs.

\section{Summary}\label{summary}

In summary, we studied the formation of BICs poisoned by MBSs in a system composed by a triple QD array with side-coupled TSCs nanowires. The BICs developed topological characteristics due to the presence of the MBSs built in the hybrid structure. Asymmetry in QDs energy levels as well as tuning of inter MBSs coupling, driving TSCs from long to short wire limit (or vice versa) using applied gate voltages \cite{Alicea2011}, can control the appearance of BICs poisoned by MBSs. We also have shown that this behavior is not restricted to a specific phase difference between both TSCs. Our findings can be seen as a way to implement protection of one of the BICs with non-topological characteristics due to the presence of MBSs.

\ack The authors acknowledge financial support from FONDECYT under Grants 1140571, 1151316 $\&$ 1140388 and from CONICYT under Grant PAI-79140064 and Act-1204. J.P.R.-A is grateful for the funding of CONICYT-Chile scholarship No. 21141034 and PIIC-UTFSM grant.

\section*{References}
\bibliographystyle{iopart-num}
\bibliography{biblio}

\end{document}